\documentclass[11pt, a4paper]{article}
\pdfoutput=1

\usepackage{comment}
\usepackage{amsmath}
\usepackage{amsfonts}
\usepackage{amssymb}
\usepackage{graphicx}
\usepackage{mathrsfs}
\usepackage{latexsym}
\usepackage{color}
\usepackage{slashed, cancel}
\usepackage{hyperref}
\usepackage{bbold}
\usepackage{soul} 

\usepackage[font=footnotesize,labelfont=bf]{caption} 

\usepackage{enumitem}

\usepackage{journals}

\usepackage[square, numbers, comma, sort&compress]{natbib} 
\hypersetup{urlcolor=blue, colorlinks=true} 

\usepackage{fancyhdr}
\numberwithin{equation}{section}

\definecolor{rossos}{rgb}{0.8,0.2,0.3}
\definecolor{bluscuro}{rgb}{0.15, 0.2, .85}
\definecolor{bluchiaro}{cmyk}{1,.3,0.,0.1}
\hypersetup{colorlinks, citecolor=bluscuro, linkcolor=bluscuro, urlcolor=bluscuro}

\makeatother   

\newcommand{\TeV}{{\rm TeV}}

 \def\be   {\begin{equation}}   \def\ee   {\end{equation}}
 \def\ba   {\begin{array}}      \def\ea   {\end{array}}
 \def\bea  {\begin{eqnarray}}   \def\eea  {\end{eqnarray}}
 \def\bean {\begin{eqnarray*}}  \def\eean {\end{eqnarray*}}

\begin{document}

\begin{flushright} 

\end{flushright}

\vspace{0.5cm}
\begin{center}

{\LARGE \textbf {
Large Extra Dimensions at LHC Run 2
}}
\\ [1.5cm]

{\large
\textsc{Giorgio Busoni}$^{\rm a,}$\footnote{\texttt{giorgio.busoni@unimelb.edu.au}}
}
\\[1cm]

\large{
$^{\rm a}$ 
\textit{University of Melbourne and CoEPP, Melbourne Node}
}
\end{center}

\vspace{0.5cm}

\begin{center}
\textbf{Abstract}
\begin{quote}

I extract new limits on the coefficient of the EFT
operator generated by graviton exchange at tree-level the ADD model 
from $pp \rightarrow jj$ angular distributions at LHC: $M_T > 6.8\TeV$ (CMS after $2.6 fb^{-1}$ of integrated luminosity) and $M_T > 8.3\TeV$ (ATLAS after
$3.6 fb^{-1}$). I also compare such limits to the ones obtained using the full graviton amplitude, and discuss the impact of additional constrains arising from other datasets, such as Mono-Jet.

\end{quote}
\end{center}

\def\thefootnote{\arabic{footnote}}
\setcounter{footnote}{0}
\pagestyle{empty}

\newpage
\pagestyle{plain}
\setcounter{page}{1}

\section{Introduction}

The new data from Run 2 of LHC, thanks to its center of mass energy increased up to $13 \TeV$, can significantly increase many limits on BSM physics. In this paper I consider the ADD (Arkani-Hamed, Dimopoulos, Dvali) model \citep{ArkaniHamed:1998rs,ArkaniHamed:1998nn, Antoniadis:1998ig}. Standard Model fields can only propagate in a $(3+1)$-D brane, while gravity is free to propagate in the  D-dimensional space ($D = 4 + \delta$), the $\delta$ extra dimensions are flat and compact. This scenario can lower the energy scale of quantum gravity from the plank scale to a much lower energy scale $M_D$, and the high value of the Plank scale is explained by the fact that the extra dimensions are large:
\be
G_N \sim \frac{1}{M_{Planck}^2} \sim \frac{1}{M_D^{2+\delta}R^\delta}
\ee
Therefore this is one of the possible solutions to the hierarchy problem. It is possible to make some predictions for collider experiments even without knowing the full UV quantum gravity theory, in the low energy limit \citep{Giudice:1998ck}.\\
One can identify different kind of signals that could be observed at LHC \citep{Giudice:2004mg,Franceschini:2011wr,Contino:2001nj}:
\begin{enumerate}
\item \textbf{Tree-level exchange of virtual gravitons} generating an EFT operator $\tau$ of dimension $8$:
\end{enumerate}
\be
{\cal L}_{int} = C_{\tau} \times \tau = \frac{8}{M_T^4} \times \frac{1}{2} \Big( T^{\mu \nu} T_{\mu \nu} - \frac{T_\mu^\mu T_\nu^\nu}{\delta-2}\Big)
\label{eq:virtualop}
\ee
where $T_{\mu \nu}$ is the energy-momentum tensor of Standard Model particles. As discussed in \citep{Giudice:1998ck}, this operator is predominantly generated by the ultraviolet part of the graviton spectrum (except in the case  $\delta = 1$, that is usually not considered, as the additional dimension would be too large and would modify gravity on astrophysical scales, even though there are some possible ways out \citep{Giudice:2004mg}). For this reason the parameter $M_T$ cannot be determined without knowing the UV quantum gravity theory, but is is possible to parameterize it as a function of some cutoff energy scale $\Lambda$. This operator contributes to parton-level processes $q\bar{q},gg \rightarrow f\bar{f},gg,\gamma\gamma$, and can therefore be probed using Dijet, Dilepton and Diphoton.
\begin{enumerate}[resume]
\item \textbf{Missing $p_T$ from emission of real massive gravitons} of the Kaluza-Klein tower. This signal is independent of the ultraviolet cutoff $\Lambda_{eff}$ as long as the collider energy is higher than such cutoff. This operator contributes to parton-level processes $qg \rightarrow qg,q\gamma$, and can be probed therefore with Mono-Jet and Mono-Photon. For more details check \citep{Giudice:1998ck,Giudice:2004mg,Giudice:2003tu}.
\item \textbf{Virtual graviton exchanges at one-loop level}, that may become more important than the tree-level operator $\tau$ of Eq.\ref{eq:virtualop} because they generate EFT operators of dimension 6, while the tree-level exchange generates the dimension-8 $\tau$ operator \citep{Giudice:2003tu}.
\end{enumerate}
In its first stage of Run 2 with lower statistics, LHC is more sensitive to the operator in Eq. \ref{eq:virtualop}, that gets enhanced due to its high dimensionality thanks to the high center of mass energy of the collisions at LHC. Limits coming from Monojet are considerably weaker using the current datasets because the statistics are not yet high enough to consider values for the $p_T$ cuts that would yield an observable signal to background ratio for values of $M_D$ yet unconstrained by Di-jet. Because of this, in this work I will derive limits on the parameters of the model only considering the Virtual Graviton Exchange, using Di-jet angular distributions data \citep{CMS:2015djr,ATLAS:2015nsi}. Such limits are usually stronger \citep{Franceschini:2011wr} than the ones coming from Di-lepton or Di-photon.\\
In section \ref{sec:virtual} I summarize how $M_T$ can be linked to $M_D$ and $\delta$. In section \ref{sec:approx} I will show that the present data can already set on $M_T$ bounds that are stronger than any other obtained by previous experiments, as summarized in table \ref{tab:mtlimits}.
In section \ref{sec:full} I will show the exclusion plots resulting from the fit with the full amplitude of section \ref{sec:approx}. Section \ref{sec:conclusions} contains the conclusions.
\section{Tree level Virtual Graviton Exchange Effective Operator}
\label{sec:virtual}
In the ADD model in the low energy approximation, one can reconsider Eq. \ref{eq:virtualop} and calculate the exact coefficient for the operator $\tau$. This, in general, is a function of the center of mass energy $s$, that is usually called ${\cal S}(s)$ \citep{Franceschini:2011wr}, and is obtained by summing over the tower of Kaluza-Klein (KK) gravitons. This sum can be usually approximated with an integral \citep{Franceschini:2011wr} over $q$, the graviton momentum. This integral is UV divergent for $\delta > 1$, and can be regularized by imposing an arbitrary cutoff $\Lambda$. The resulting operator gets a coefficient that depends on the center of mass energy \citep{Hewett:1998sn}
\bea
\tau = {\cal S}(s) \left(T_{\mu \nu} T^{\mu \nu} - \frac{T^\mu_\mu T^\nu_\nu}{\delta + 2} \right)
\label{eq:fullamp}
\eea
The value of the ratio $\Lambda/M_D$ gives the effective coupling of gravity, so larger (lower) value will mean the gravity is strongly (weakly) coupled \citep{Giudice:2003tu}. 
For $\delta > 2$ the main contribution to the integral comes from the heaviest graviton with mass $m \sim \Lambda$ and thus, if $s \ll \Lambda^2$, one can consider the $s\rightarrow0$ limit for the function $S$, in this way the scattering amplitude can be estimated by the EFT operator of Eq. \ref{eq:virtualop} with a coefficient that is usually defined as $8/M_T^4$ (for other conventions, see \citep{Hewett:1998sn}).

However, as the dominant LHC bound will come from the most energetic events because of the high dimension of the operator (considering only high-energy regions also helps reducing the SM QCD background), it is more appropriate to keep the full amplitude, including the $\Lambda$ dependence.

\section{Fit with Run 2 Data}
\label{sec:fit}
To fit the data one needs to calculate the expected signal as a function of the parameters of the model. In this case, the signal  corresponds to the binned angular distributions of the jets as a function of the variable $\chi$.
\be
\chi = e^{|y_1-y_2|}
\ee
Where $y_1,y_2$ are the rapidities of the 2 jets. 
The SM expects such distributions to be almost flat, as QCD cross sections in SM are predominantly coulomb-like. On the other hand, interactions originated by the operator of Eq. (\ref{eq:virtualop}), have a different angular distribution, more enhanced at high $p_T$, as they are mediated by a spin-2 particle, that therefore differs significantly from QCD and would imply a deviation from a flat distribution, peaking at low values of $\chi$.
The bins where choosen in the same way as in \cite{ATLAS:2015nsi} and \cite{CMS:2015djr}. The expected signal was calculated on a grid of the parameters $M_D,\Lambda/M_D$, then interpolated. Using the interpolated grid I could compare data with the expected theoretical value and find the $95\%$ CL bounds on the parameters of the theory (either $M_T$ or $M_D,\Lambda$) by imposing
\be
\chi^2 = \sum_{i}^{\rm bins} \frac{(t_i - \mu_i)^2 }{ \sigma_{i ~\rm stat}^2 + \sigma_{\rm syst}^2} < \chi^2_{\rm min} + 3.84
\ee
where $\mu_i$ are the experimental values, $\sigma_{i ~\rm stat}$ are the statistical errors, $\sigma_{\rm syst} \approx 0.03$ is an estimate of the systematic uncertainties and $t_i$ are the theoretical predictions obtained from the interpolated grid.
For the calculation of the expected signal, I used formul\ae{} for the cross sections for tree-level graviton effects taken from the appendix of \citep{Giudice:2004mg}, and I implemented them in a ad-hoc Mathematica integration package. I had previously verified that using this approach the various distributions that one obtains correctly reproduce the ones obtained with {\sc MadGraph} \citep{Alwall:2011uj} and {\sc PYTHIA8} \citep{Sjostrand:2007gs}. Hadronization and jet reconstruction should deliver negligible effects \citep{Franceschini:2011wr} and therefore I neglected them in this analysis.
I implemented the same cuts as CMS
\be
\chi < 16, |y_{boost}| < 1.11
\ee
and ATLAS
\be
\chi < 30, |y_{boost}| < 1.1
\ee
The analysis was repeated using different PDF sets MSTW2008 \citep{Martin:2009iq}, MMHT2014 \citep{Harland-Lang:2014zoa}, CT6 \citep{Nadolsky:2008zw}, CT14 \citep{Dulat:2015mca} and NNPDF 2.3 \citep{Ball:2012cx} and NNPDF 3.0 \citep{Ball:2014uwa}, all at NLO. The results do not depend significantly on the PDF set used, therefore I will just show the results obtained using the CT14PDF set.

The angular distributions in QCD are sensible to NLO effects. This sensitivity can be reduced by kinematical cuts, but it can still have important effects at very high invariant masses \citep{Boelaert:2010kba}. Therefore, for a more rigorous analysis, it is necessary to apply QCD NLO corrections to the expected signal in order to better estimate the correct limits on the model. Very high invariant masses make also EW corrections have a sizable effect \citep{Dittmaier:2012kx} on both for the shape and normalization of the distributions. To take both these effects into account, K factors where applied to the pure QCD signal.

\subsection{Approximated effective operator}
\label{sec:approx}

\begin{figure}[t!]
\centering
\includegraphics[width=0.6\textwidth]{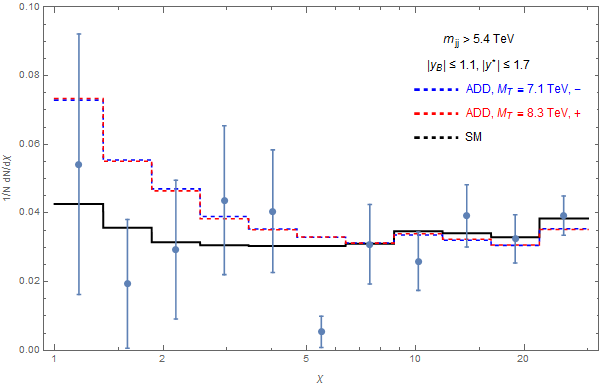}
\caption{\em $pp\to jj$ angular distribution with $M_{jj}>5.4\TeV$ in the SM (black line) and for ADD with $M_T = 8.3\TeV$ (positive interference, red dashed line) and $M_T =7.1\TeV$ (negative interference, blue dashed line).
The data are from ATLAS \cite{ATLAS:2015nsi}.
\label{fig:distributions}}
\end{figure}

I compare the first Run II data \citep{CMS:2015djr} and \citep{ATLAS:2015nsi} to the new physics described by Eq. \ref{eq:virtualop}. The following analysis applies to any number $\delta > 2$ of extra dimensions, as the double trace term in ${\cal T}$ having a $\delta$-dependent coefficient is irrelevant because the masses of the particles involved in the collision are much smaller than the center of mass energy.

In Fig \ref{fig:distributions} I show the SM predictions for the angular distributions for ATLAS kinematical cuts \citep{ATLAS:2015nsi}. Together with them I show the distributions for $M_T = 8.3 \TeV$ ($M_T = 7.1 \TeV$), for positive (negative) interference.

\begin{figure}[ht!]
\centering
\includegraphics[width=0.49\textwidth]{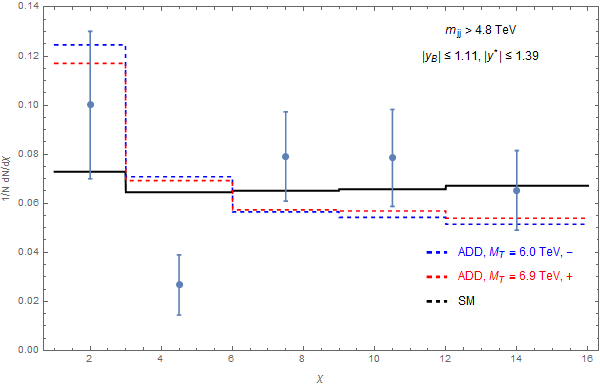}
\hspace{-0.1cm}
\includegraphics[width=0.49\textwidth]{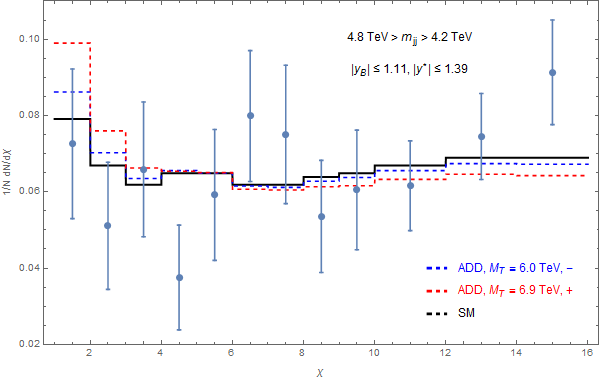}
\caption{\em $pp\to jj$ angular distribution with $4.8\TeV>M_{jj}>4.2\TeV$ (left panel) and $M_{jj}>4.8\TeV$ (right panel) in the SM (black line) and for ADD with $M_T = 6.8\TeV$ (positive interference, red dashed line) and $M_T = 6.0\TeV$ (negative interference, blue dashed line).
The data are from ATLAS \cite{ATLAS:2015nsi}.
\label{fig:distributions2}}
\end{figure}

In Fig \ref{fig:distributions2} is the analogue of Fig \ref{fig:distributions} for CMS kinematical cuts \citep{CMS:2015djr}. The chosen values of $M_T$ in this case are $M_T = 6.8 \TeV$ ($M_T = 6.0 \TeV$), for positive(negative) interference. The two panels refer to the two most energetic signal regions.

\begin{table}[ht!]
\centering
\begin{tabular}{|c|cc|}
\hline\hline
Experiment& + & $-$\\ 
\hline
ATLAS at 7 TeV with 36/pb & 4.2 & 3.2\\
CMS at 7 TeV with 36/pb & 4.2&3.4\\
CMS at 13 TeV with 2.6/fb & 6.8& 6.0\\
ATLAS at 13 TeV with 3.6/fb & 8.3& 7.1\\
\hline
Projected sensitivity Run 2 & 9.5 & 7.5\\
\hline\hline
\end{tabular}
\caption{{\bf Tree-level graviton exchange}:
\em { 95\% CL bounds on $M_T$ (in {\rm TeV}) for the dimension-8 operator of Eq.\ref{eq:virtualop}
for positive and negative interference, and projected sensitivity for Run 2 with $3000 fb^{-1}$ integrated luminosity and systematics of $1\%$.
\label{tab:mtlimits}}
}
%
\end{table}

The obtained limits on $M_T$ are reported in Tab. \ref{tab:mtlimits}, together with the ones \citep{Franceschini:2011wr} coming from older datasets, for comparison purposes, and the projected exclusion reach for Run 2 with $3000 fb^{-1}$ integrated luminosity and systematics of $1\%$. The limit on $M_T$ coming from CMS dataset obtained in this simplified analysis reproduces the one from CMS preliminary analysis \citep{CMS:2015djr} up to 15\%. This difference is probably due to different QCD NLO K factors that significantly affects the bounds, as reported previously.

The result of the analysis indicates that the limits obtained from ATLAS are even higher than the ones from CMS. This is probably due to the higher integrated luminosity reached by ATLAS, that permits the use of a more stringent kinematical cut on the invariant mass of the two jets for the most energetic signal region.

\subsection{Full graviton exchange amplitude}
\label{sec:full}

\begin{figure}[ht!]
\centering
\includegraphics[width=0.38\textwidth]{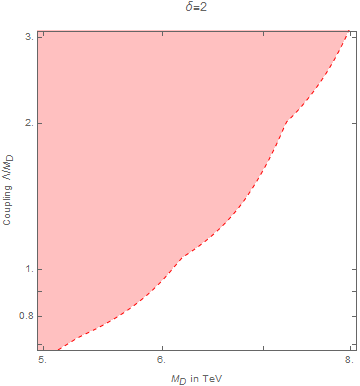}\hspace{0.5cm}
\includegraphics[width=0.38\textwidth]{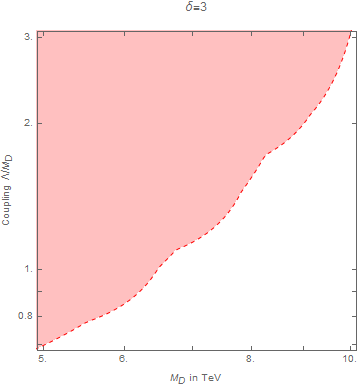}\\
\includegraphics[width=0.38\textwidth]{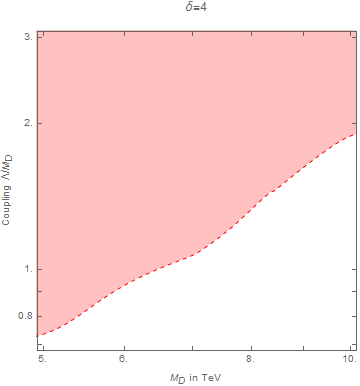}\hspace{0.5cm}
\includegraphics[width=0.38\textwidth]{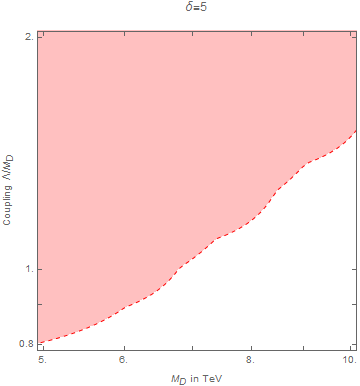}\\
\includegraphics[width=0.38\textwidth]{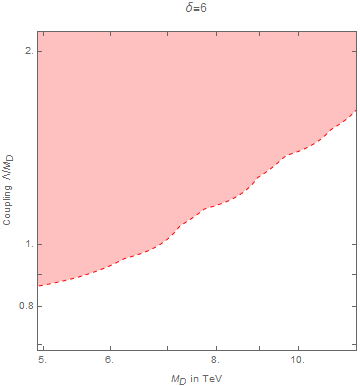}
\caption{\em
The colored area is the region excluded at 95\% CL by 
ATLAS after {\rm 3.6/fb}.
\label{fig:exclusionplots}}
\end{figure}

I now use the operator of Eq. \ref{eq:fullamp} to set limits in the $M_D,\Lambda/M_D$ plane. Using the energy-dependent coefficient ${\cal S}(s)$ one expects to find stronger limits, as the coefficient grows with the center of mass energy. The results of the fit are shown in Fig.\ref{fig:exclusionplots}. The $95\%$ CL bounds are defined as $\chi^2< \chi_{SM}^2 + 3.84$, as before. The ratio $\Lambda/M_D$, as said before, is an effective parameter of the unknown strength of quantum-gravity.
%

\section{Conclusions}
\label{sec:conclusions}

Using the very first dijet datasets from CMS and ATLAS $pp \rightarrow jj$, despite the the uncertainties intrinsic of the hadronic nature of the final states and the the fact that the statistics is still low, new limits on the coefficient of the dimension-8 operator $\tau$ were derived. This is thanks to the high dimension of the operator and the new high energy delivered by LHC.

In the second part of the work I went beyond the EFT approximation and used the full amplitude produced by graviton exchange at tree level in terms of an effective cut-off parameter $\lambda$, that is the mass of the heaviest KK graviton. Fig. \ref{fig:exclusionplots} shows the resulting bounds from in the $(MD,\Lambda/MD)$ plane.

\vspace*{50px}

\textbf{Acknowledgements} I thank Alessandro Strumia and Andrea De Simone for interesting discussions and comments on the manuscript.


\newpage

\label{Bibliography}

\lhead{\emph{Bibliography}} 

\bibliography{Bibliography} 

\end{document}